# Acceleration and energy consumption optimization in cascading classifiers for face detection on low-cost ARM big.LITTLE asymmetric architectures

A. Corpas [1], L. Costero [2], G. Botella [2], F. D. Igual [2], C. García [2], M. Rodríguez [1]

[1] Dept. of Architecture and Computer Technology, ETSIIT, University of Granada, 18071, Granada, Spain.
{albertocorpas@correo.ugr.es, manolo@ugr.es}

[2] Dept. Computer Architecture and Automation, Complutense University of Madrid, 28040, Madrid, Spain.
{lcostero, gbotella, figual, garsanca} @ucm.es

**Abstract:**

This paper proposes a mechanism to accelerate and optimize the energy consumption of a face detection software based on Haar-like cascading classifiers, taking advantage of the features of low-cost Asymmetric Multicore Processors (AMPs) with limited power budget. A modelling and task scheduling/allocation is proposed in order to efficiently make use of the existing features on big.LITTLE ARM processors, including: (I) source-code adaptation for parallel computing, which enables code acceleration by applying the OmpSs programming model, a task-based programming model that handles data-dependencies between tasks in a transparent fashion; (II) different OmpSs task allocation policies which take into account the processor asymmetry and can dynamically set processing resources in a more efficient way based on their particular features.
The proposed mechanism can be efficiently applied to take advantage of the processing elements existing on low-cost and low-energy multi-core embedded devices executing object detection algorithms based on cascading classifiers. Although these classifiers yield the best results for detection algorithms in the field of computer vision, their high computational requirements prevent them from being used on these devices under real-time requirements. Finally, we compare the energy efficiency of a heterogeneous architecture based on asymmetric multicore processors with a suitable task scheduling, with that of a homogeneous symmetric architecture.

## 1.- Introduction

Face detection [1] constitutes a significant part of computer vision, and is especially related to communication and human-computer interaction. However, given that the human face is a dynamic object with a high degree of variability in its appearance, its detection becomes a challenging task to deal with in computer vision. Due to the complexity of the algorithms related to face detection, a large amount of computational resources and memory is required. Hence, the software implementations of these algorithms become quite inefficient when they are required to be executed on low-cost and low-energy embedded systems due to their limited resources and low performance. In these cases, optimization techniques based on software parallelization can be applied to accelerate the parts which require more computational resources in detection processes.

In this context, the most successful algorithms for face detection are usually executed on real-time systems on high-end CPUs in order to leverage their high processing power [2,3]. However, similar implementations executed on low-power CPUs (e.g. those present on mobile devices) will not work fast enough to meet real-time restrictions. This gap in performance is caused by the diversity of features existing on the processors used in mobile devices, which are optimized to be low-cost and provide low energy consumption, and therefore their performance is behind that of server-level processors.

Currently, mobile devices are evolving from single-core CPUs to multi-core CPUs, following a similar progression to that observed in general-purpose architectures over the last decade. As of today, smartphones and handheld devices commonly feature multi-core processors with up to eight processing cores, and there is no doubt that the number will increase in forthcoming products. This same trend applies to embedded microcontrollers.

This trend implies changes in the way software developers deal with performance improvement. Thus, improving the execution performance of a sequential application, initially written to be executed on traditional single-core CPUs, implies dramatic changes in the implementations to be able to exploit the potential parallelism of multi-core CPUs. In this sense, the OpenMP API [4] is one of the best options in parallel programming for shared memory architectures, since it is supported on different operating systems, compilers and hardware devices, even being able to work on mobile devices nowadays.

The OpenMP standard supports task-based parallelism since its third release. This functionality relies on task and data dependencies annotations in the source code provided by the developer, and then exploited at runtime by a task scheduler to exploit out-of-order execution without user's intervention. These kind of extensions were inspired, among others, by the efforts introduced in OmpSs [5], which is a portable and scalable programming model that provides programmers with an easy and flexible interface to develop parallel applications with minimal code modifications from sequential implementations. The OmpSs API uses a programming paradigm based on directives which make it possible to expose parallelism of an already existing sequential program. Once the application is compiled with Mercurium [6], which is the OmpSs compiler, it can be executed through Nanox [7], the OmpSs runtime, which consists of a set of libraries in charge of controlling the program’s execution and making sure this is completed in the most efficient way.

In asymmetric multi-core architectures, the so-called big.LITTLE [8] ARM processors are especially relevant. Big.LITTLE processors include powerful cores (big) together with other low-energy and low-performance cores (LITTLE), both sharing the same instruction set (ISA). In this work, the chosen asymmetric architecture is the Odroid XU4 [9, 10, 11], as shown Figure 1. This board consists of a Samsung Exynos 5422 SoC (System-On-Chip) built on 28nm, which includes an 8-core big.LITTLE ARM processor. The eight cores are grouped into two clusters with four cores in each one; the big cluster features 4 high-performance Cortex A15 cores, while the LITTLE cluster includes low-power Cortex A7 cores. For comparison purposes, we also use a Raspberry PI 3 B+ [12,13]; this board features a 4-core CPU based on ARM Cortex-A53 and represents a good example of an affordable embedded device which, together with the Odroid XU4, is comparable to CPUs existing in current smartphones (Android, iPhone and Windows Mobile) since their processors also include several ARM cores [14]. Furthermore, any improvement made to this platform is easily portable to smartphones and tablets.

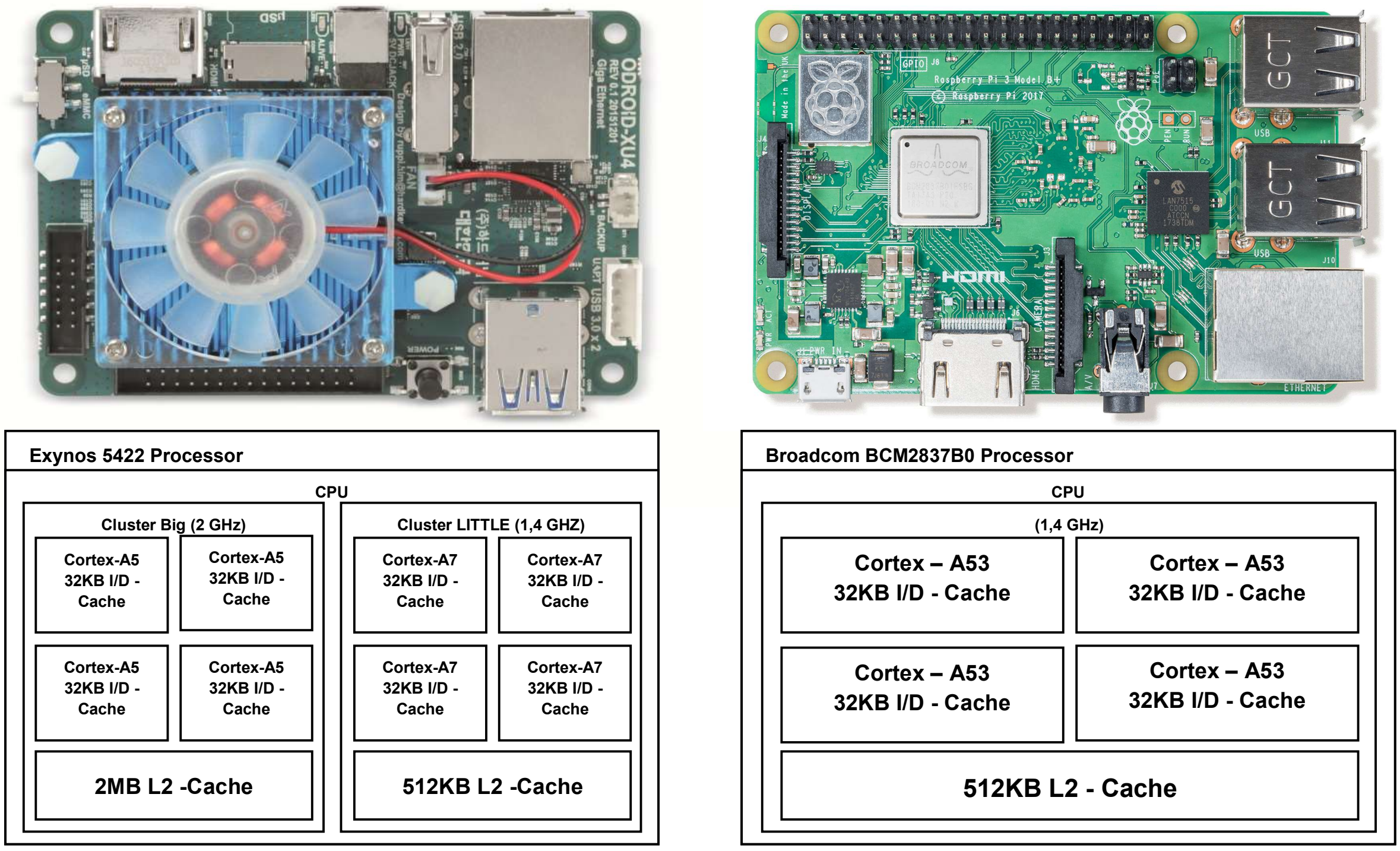

Exynos 5422 hosting eight processors, 4 cluster Big and 4 cluster LITTLE

Broadcom BCM2837B0 quad-core A53 (ARMv8) 64-bit 1.4GHz

***Figure 1:** Odroid XU4 (left) and Raspberry Pi 3 Model B+ (right) boards.*

## 2.- Method used and testing environment

Face detection consists in determining whether there is a face in an arbitrary image, and if this is the case, identifying the position of the face. Therefore, a reliable face detector must be able to find all the existing faces in an image. The traditional methods for face recognition can be split into two groups: (I) *holistic methods*, which are based on image correlation and use comparison models for the recognition process; (II) *geometric methods*, which compare different geometric characteristics of the faces. Thus, there are several algorithms for face detection, each one based on any of the known techniques for this purpose, such as neural networks, closest neighbor, etc. Among them, one of the most frequently implemented and with the greater advantages is the algorithm proposed in 2003 by Viola-Jones [15]. This method was the first one to offer robust detection in real time, allowing fast image processing and a high detection rate. It is also worth mentioning that the algorithm can also be trained to detect any kind of object. In this context, the Viola-Jones algorithm [15] has been chosen for its implementation in the proposed system for face detection. For this purpose, a C++ simplified implementation of the algorithm will be used, which has fixed training parameters and provides a high detection rate for a broad range of input images.

Our main goal is to adapt, evaluate and tune a sequential C++ implementation targeting asymmetric architectures. At this point, the steps followed to accomplish this goal are:

1. Initial measurement of the execution time of the application in sequential mode, and analysis of the results through the profiling tool Gperftools, provided by Google which works using time-based sampling, which allows an accurate analysis of applications based on multiple sub-processes. For this aforementioned reason, it is a suitable tool for collecting the necessary information for the latter software acceleration step using OmpSs.

2. Parallelism extraction at task level using the OmpSs programming model [5]. For this, a directed acyclic graph (DAG) is created in order to represent the task parallelism existing in the algorithm, which will allow the optimization of the computational resources.

3. Development of techniques for energy consumption reduction and to exploit, in an efficient way, the computational resources offered by the asymmetric architectures [16,17].

4. Measurement of energy consumption in the testing environment created for the ODROID XU4 and Raspberry Pi 3 B+ boards.

5. Experimental analysis of the attained results.

### 3.- Viola-Jones face detection algorithm

The Viola-Jones algorithm [15] for face detection consists of two main stages: a *training phase*, and the actual *detection phase*. The training stage is based on the AdaBoost [15] boosting algorithm, and it is the most time-consuming part. In the second stage, the previously trained detector is applied to each of the images to be analyzed. This phase is faster, and can be executed in real-time to detect the trained objects.

In this study, the starting point is a trained system for face detection where the size and location of the features that will indicate the existence of a face inside a detection window is already known. To obtain this information, an exhaustive exploration of the image is performed, evaluating features in distinct positions and scales in order to take into account different face sizes. This results in a large volume of data processing for each analyzed image.

The Viola-Jones detection method uses groups of simple and common face features. The use of these features leads to a higher speed at detecting faces than methods based on pixels. The features used to detect faces are similar to the Haar-like ones proposed by Papageorgiou et al. [18]. Viola-Jones uses five types of features in its detection system, two of them based on two rectangles, two based on three rectangles and another one based on four rectangles, as can be seen in Figure 2.

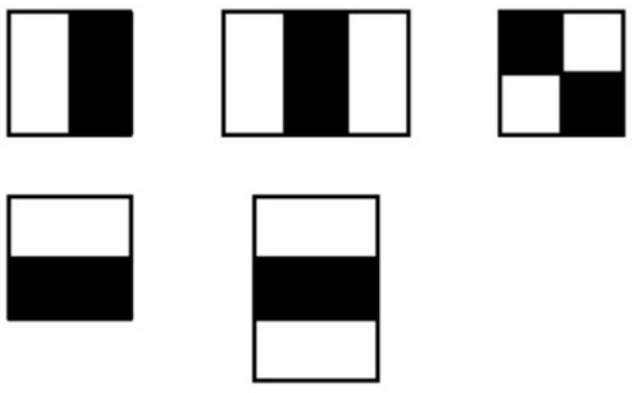

***Figure 2:*** *Haar-like features of 2, 3 and 4 rectangles as defined by Viola-Jones work.*

The features are not only characterized by their form, but also by their size and position inside the detection window, as well as their contribution to face detection. For this reason, it is necessary to calculate the value of the feature, that is, the difference between the intensities of the points in the white area and the intensities in the black area as shown in Figure 2. If the value of a feature is above a specific threshold (classification threshold), it is considered that the feature contributes with a specific 'alpha' value to the detection of a face.

Thus, the scalar value of a feature is obtained by the sum of the pixel values of each rectangle according to the following formula:

$$Feature_j = \sum_{1 \le i \le N} w_i \cdot rectangular_sum(r_i) \qquad (1) ,$$

where $\{r_1, \ldots, r_N\}$ is the vector of rectangles that constitute the feature, and $w_i$ is the weight of each one. Furthermore, *rectangular_sum* indicates the addition of the pixel value. Viola-Jones establishes a minimum detection window for an image of 24x24 pixels. In a window of these characteristics it is possible to find up to 45,396 different features (a number that is higher than the number of pixels in the window). However, it is not necessary calculate the value of every single feature in order to detect a face, as a small portion of these features, when adequately selected, can be enough to discern, with a low error rate, whether a region is a face or not.

In order to select the best features to classify faces, we use a boosting algorithm called AdaBoost, which is based on the combination of several simple or weak classifiers (in this case, based on Haar-like features, see Figure 2). This combination creates a more complex classifier (strong classifier) with a lower error rate than each of the individual weak classifiers. This means that each weak classifier is related to a single feature and the combination of all the weak classifiers creates a strong one. Formally, a classifier based on a single feature j is represented as in Formula 2.

$$h_j(x) = \begin{cases} 1 & if\ p_j f_j(x) < p_j \theta_j \\ 0 & Otherwise \end{cases} \qquad (2)$$

Where $f_j(x)$ is the value obtained when feature j is applied over image x, $\theta_j$ is the threshold of the feature and $p_j$ is its polarity. This last value could be 1 or −1 and allows the inversion of the feature, turning positive rectangles into negative ones, and vice versa.

Next, the learning algorithm Adaboost is defined both for the selection of characteristics and for the training of each one of the stages of the cascade classifier [15].

- Given a set of images $(x_1, y_1),\ldots,(x_n, y_n)$ where $y_i = 0, 1$ for negative and positive examples respectively.
- Initialize weights $w_i = \frac{1}{2m}, \frac{1}{2l}$ *for $y_i$= 0, 1 respectively, where m and l are the number of negatives and positives respectively.*
- *For t= 1,…,T:*

1. Normalize the weights, $w_{t,i} \leftarrow \frac{w_{t,i}}{\sum_{j=1}^{n} w_{t,j}}$
2. Select the best weak classifier w.r.t the weighted error

$$\epsilon_t = min_{f,p,\theta} \sum_i w_i \left| h(x_i, f, p, \theta) - y_i \right|$$

3. Define ht(x) = h(x, ft, pt, θt) where ft, pt, and θt are the minimizer of ϵt.
4. Update the weights

$$w_{t+1,i} = w_{t,i} \beta_t^{1-e_i}$$

Where $e_i$ = 0 if example $x_i$ is classified correctly, $e_i$ = 1 otherwise, and $\beta_t = \frac{\epsilon_t}{1-\epsilon_t}$.

5. The final strong classifier becomes:

$$c(x) = \begin{cases} 1 & \sum_{t=1}^{T} \alpha_t h_t(x) \geq \frac{1}{2} \sum_{t=1}^{T} \alpha_t \\ 0 & otherwise \end{cases}$$

Where $\alpha_t = \log \frac{1}{\beta_t}$

**Figure 3:** Adaboost learning algorithm. T hypotheses are constructed each one using a single feature. The final hypothesis is a weighted linear combination of the T hypotheses where the weights are inversely proportional to the training errors.

The feature calculation would be a costly computational process if it was not possible to calculate the integral image. The value of all the points inside any rectangle in the image can be calculated quickly. The integral image facilitates the calculation of the value of any feature. The formula for the integral image is shown in Equation 3, and is obtained from the grey-scale representation of the image:

$$II(x,y) = \sum_{\substack{1 \leq x' \leq x \\ 1 \leq y' \leq y}} I(x', y') \ , 1 \leq x \leq n, 1 \leq n \leq m \qquad (3)$$

Using the integral image, any sum inside a rectangle can be calculated with four references to the table, as indicated in Figure 4.

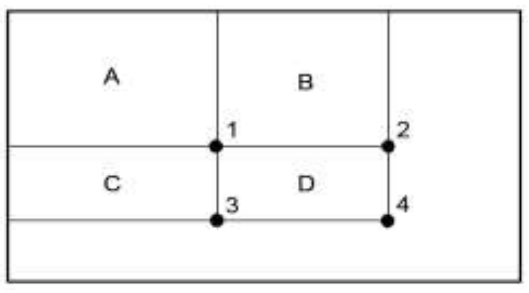

***Figure 4:*** *As an example, the sum of pixels inside rectangle D can be easily calculated as 4+1-2-3, where 4 is the result of the integral image at that point. Thanks to this representation, it is possible to calculate the sum of pixels inside a rectangle of any size by means of only 4 memory accesses.*

The only disadvantage of the integral image is that it uses four times more memory than the original image. As the integral image is the sum of the pixels of the image, it cannot be defined as a matrix of bytes as is usually done with grey-scale images, so it is necessary to use an integer matrix. This is the reason for the larger memory size, since integer type occupies 4 bytes in most systems.

The integral image makes it possible to easily calculate the value of a weak classifier. These weak classifiers are combined to build strong ones by means of AdaBoost learning algorithm and a set of selection parameters. Those parameters contain low values which make it possible to quickly discard regions with no faces in order to focus on those regions with a higher probability of containing a face. Strong classifiers designed in this way are grouped into the so-called cascading classifiers, setting up decision steps, where it is decided whether there is a face in the region or not. An example of these steps can be seen in Figure 5.

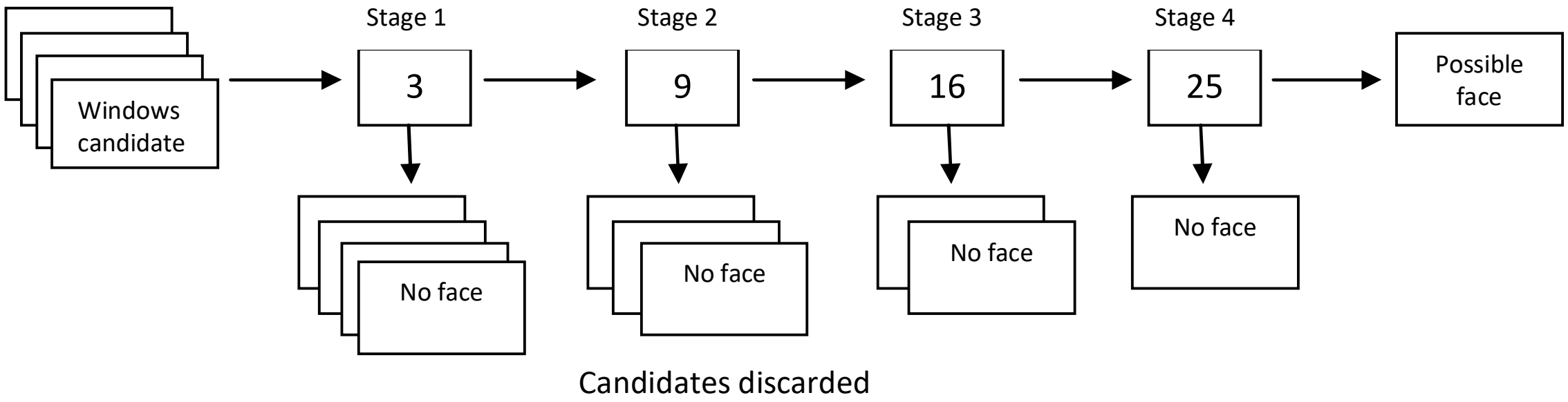

***Figure 5***: *Cascading classifiers distributed in four stages.*

In the example above, Stage 1 has a strong classifier which consists of 3 weak classifiers. This means that in this stage, three features are evaluated in the detection window. If the evaluation of these three features is above the threshold established during the training phase of the strong classifier, there is a high probability that in the region considered there is a face, and then it can be evaluated in the following stage (Stage 2 in Figure 5). In the new stage, more features are evaluated in the region, and again, if the threshold is not exceeded, the region is discarded and the algorithm evaluates another region.

The first stages have evaluated a large number of images, so the strong classifiers associated to them must be designed to have a much lower computational cost than the ones in the later stages. This leads to spending less time on areas with no faces, and more time on those which include faces. If a window is not discarded in any of the stages, the window is annotated as a possible face. The process continues sliding the window (in a horizontal or vertical direction, Figure 6) according to a specific scanning factor. Once the entire image has been considered with a specific window size, the window size or the image size increases according to a particular scaling factor, and the entry process is repeated. The values for the scanning factor and the scaling factor are also obtained during the training stage.

In addition, the face detector is applied to images of different sizes and the faces included in a particular image can also have different sizes. The system must be able to extract sub-windows from an image in order to analyze them. This is shown in Figure 6. The detector also scans an image in several locations, which means that the sub-window moves across the image by a specific number of pixels, Δ. The selection of this number will affect the detector's speed, as well as its performance. In Viola-Jones' work, the results are shown for $\Delta = 1$ and $\Delta = 1.5$, using scaling factors of 1 and 1.25, respectively.

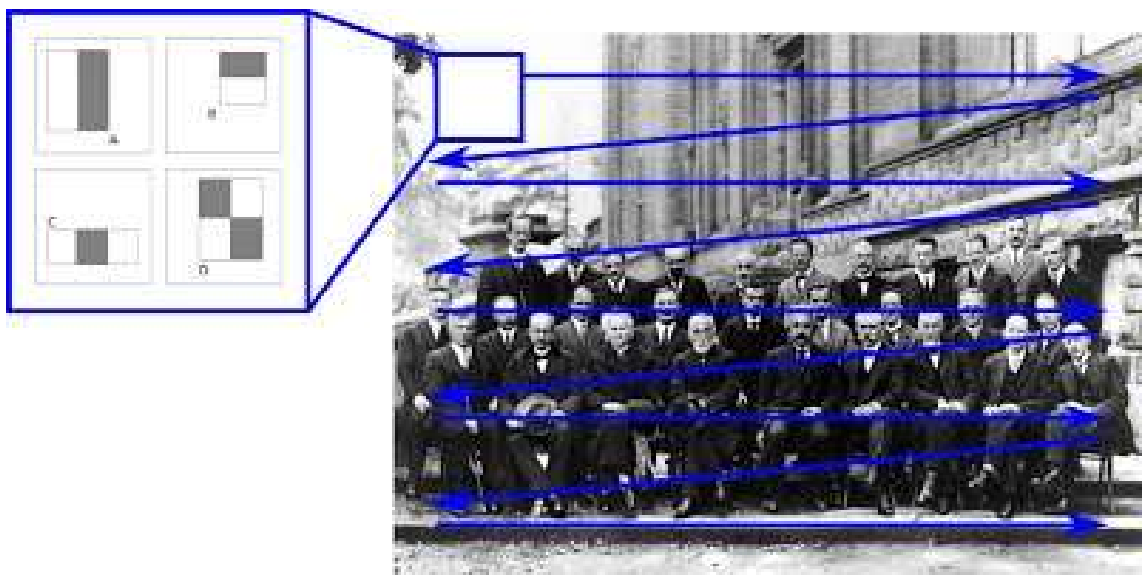

***Figure 6:*** *Slide of the detection window across the image [19].*

**4.- Sequential implementation of the Viola-Jones algorithm**

For the sequential implementation, a simplified version of the Viola-Jones algorithm has been used, specifically a version developed for low-cost embedded systems with low-power and slow processing speed requirements, together with a pre-trained file containing all the parameters involved in the cascading classifiers for face detection. This file provides a reasonable detection ratio for a large number of images.

The pre-trained file that contains the parameters of each stage of the cascade classifiers is obtained from previous training of the detection system [15]. Image sample resolution selected is 24x24 pixels.
The number of stages of the cascade classifier is obtained from the detection and performance objectives to be achieved by the system. The more features used, the higher detection rate and a lower rate of false positives will be reached up. Meanwhile, classifiers with more features will need more time to determine whether a sub-window contains a face or not. When training the classifier, the number of stages, the number of characteristics and the threshold of each stage must be optimized.

The detection rate (DR) and the false positive rate (FPR) are defined as indicated below:

$$DR = \prod_{i=1}^{k} d_i \qquad FPR = \prod_{i=1}^{k} f_i \qquad (4)$$

Where $f_i$ and $d_i$ is the FPR and the DR of stage *i* respectively, and *k* the number of stages of the cascade classifier. In this context, each stage of the cascade classifier is trained using Adaboost increasing the number of characteristics of the stage until the desired false-positive and detection rate is obtained. In this case for the system a detection rate of 95% was fixed and a low false positives rate of the order of 10-5, which is a very reasonable ratio for the system. To achieve these design indices, the number of stages of the detection system is set to 25 in total, each with a different and unique feature number.

The algorithm implementation can be carried out in two different ways depending on how faces are scaled in the image. The first way is scaling the classifiers and thus the detection window in order to cover different sizes in the image. The second consists in keeping the detection window at the same size during the entire process, scaling the image only through interpolation until a predefined minimum size is reached. In this work, the second approach is used, keeping the size of the detection window and scaling the image by different moments, building what it is known as a pyramid (Figure 7). The pyramid is a multi-scale representation of an image, so that face detection can be scale invariant, i.e. the detection of big and small faces uses the same detection window. The implementation of this pyramid is performed by reducing the image resolution by means of the algorithm based on pixel neighborhoods.

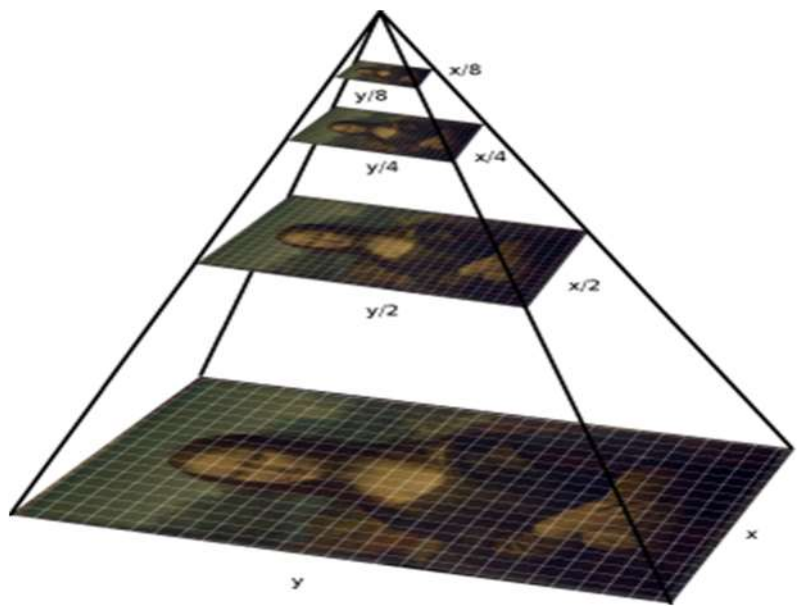

***Figure 7:*** *Pyramid. Multi-scale representation of an image [20].*

Once the image has been scaled, the integral image and the quadratic integral image (normal=$\sum x$, quadratic=$\sum x^2$) are calculated in order to obtain the statistical typical deviation. All the examples of sub-windows used in the training stage were obtained with a normalized variance, thus the impact of different light conditions is minimized. Then, the normalization through typical deviation is also necessary during image detection. The expression used in the algorithm implementation is given by the Equation 5, where N=W·H is the dimension of the detection window.

$$\left(\frac{\sigma}{N}\right)^2 = N\sum x^2 - \left(\sum x\right)^2 \qquad (5)$$

Finally, note that in the implementation a 24x24 pixel detection window and a scale factor of 1.2 have been considered. The final development consists of 25 stages, each one with a specific number of weak classifiers [21], which form the strong classifier of each stage. The first stages are the ones with lower computational complexity in order to discard the largest number of windows in the shortest time. Among all the stages there are 2913 weak classifiers, each one requiring 18 parameters previously obtained during the training process and stored in a text file. This means that to detect one single face in an image, going through any single stage of the cascading classifier, it is necessary to calculate 2913 features; each one being defined in the detection window by 18 parameters. Once the calculation process has finished, a comparison in each stage is made with the threshold values obtained during the training process.
Figure 8 addresses the algorithm pseudo-code with its main stages; meanwhile Figure 9 shows an example of face detection.

```
for each scaled image (pyramid) do
   Reduce the image scale
   Get the integral image of the current scale
   for each step of the sliding detection window do
      for each stage in the cascading classifier do
         for each feature in the stage do
            evaluate the detection window
         end
         get the accumulated value for the feature in the
             stage
         if accumulated value is below the threshold of the
             stage do
            break the loop and reject the window as a face
         end
      end
      if the detection window is above the threshold
             value for the stage do
         accept the window as a face
      else
         reject the window as a face
      end
   end
end
```

***Figure 8:*** *Pseudocode for Viola-Jones algorithm.*

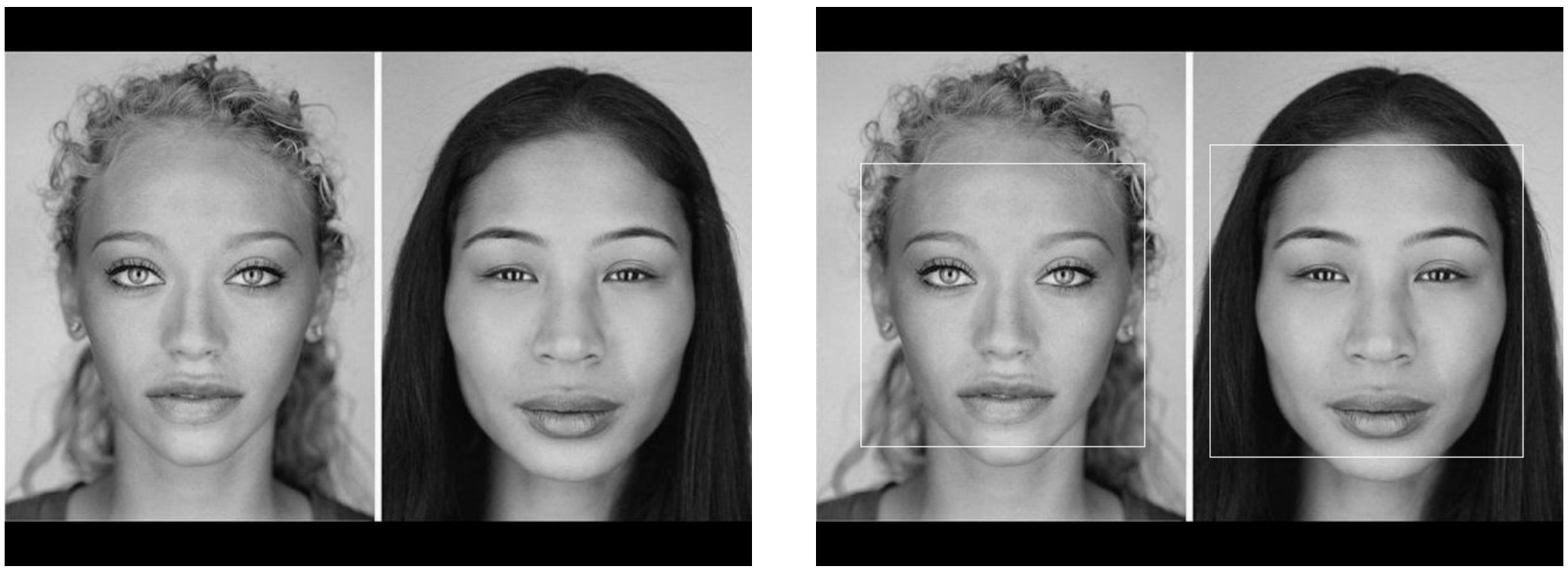

Input Output

*Figure 9: Input and output images, showing the detected faces [22].*

## 5.- Performance analysis of a sequential execution

The execution time of the software developed for face recognition clearly depends on: (I) the scale factor, set to 1.2 as optimum value; (II) the value for the slide of the recognition window across the entire image; (III) and the image resolution, since the higher the resolution is, the longer the time spent on image analysis will be.
In this context, in each one of the existing testing environments for the ODROID XU4 and Raspberry Pi boards, the execution time has been measured on a sample of 10 images with the same format and resolution, a scale factor of 1.2, the same slide factor for the detection window (1 pixel per iteration), and no software parallelization, i.e. sequential execution. Furthermore, each of the images considered has a different number and size of faces, which allows us to see the way the execution time evolves depending on the number of faces in the image. The results can be seen in Figure 10.

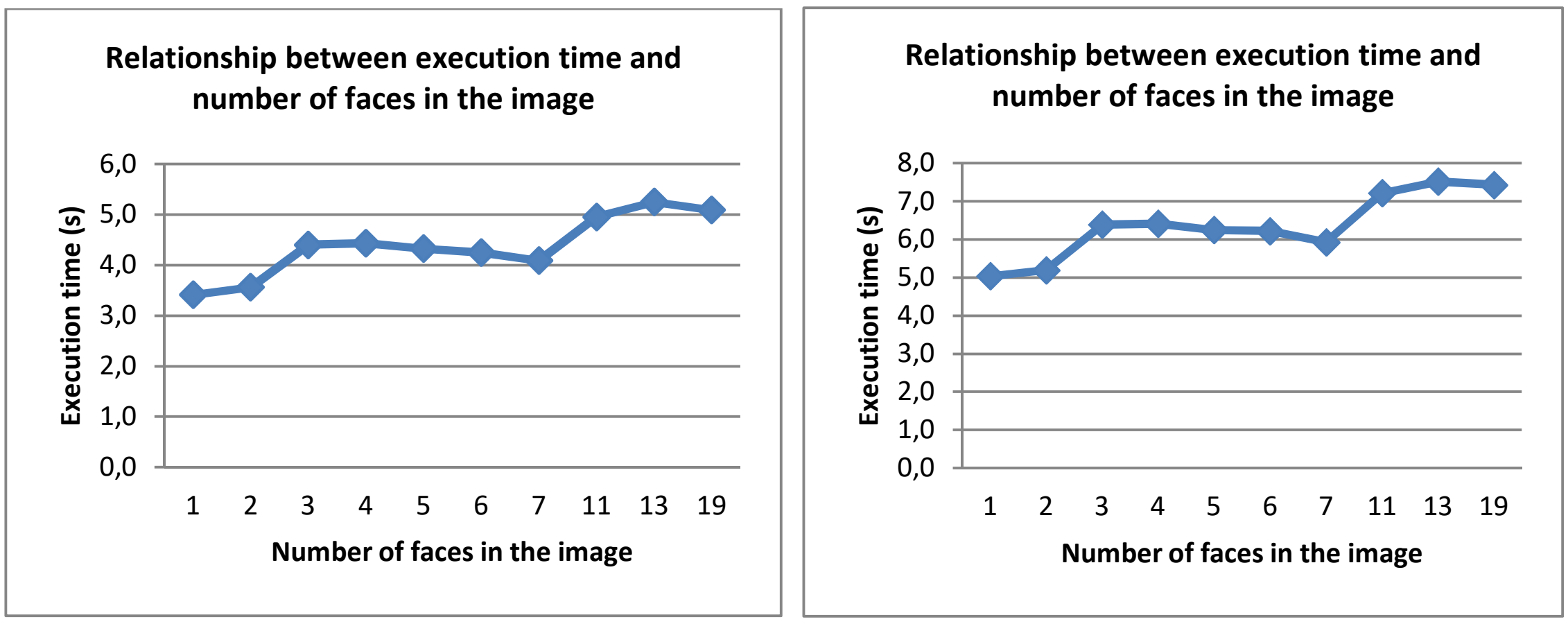

***Figure 10:** Results for **Odroid X4U** (left) and **Raspberry Pi 3 B+** (right) boards.*

From the analysis, it might be expected that the more faces there are in the image, the longer the execution time is, since each detected face needs to go through every existing stage in the cascading classifier. However, as can be seen in the figure, it does not always behave like that, which means that there must be other factors besides the number of faces that impact on the execution time and explain the variation.
After performing different tests on the sample images, a meaningful difference was detected between the images whose execution time increases when the number of faces increases, and the ones that do not follow this rule. The difference is the value of the integral image, which is higher in the ones that do not follow the rule, that is, the higher the integral image value, the shorter the execution time. The integral value is the sum of all the pixels in an image, i.e. the value of the right-most bottom pixel in the integral image of the analyzed image.
To test this observation, another sample of 10 images with the same resolution and requirements as before was chosen, but this time containing a single face per image. The results of comparing the execution time and integral value for each image are shown in Figure 11.

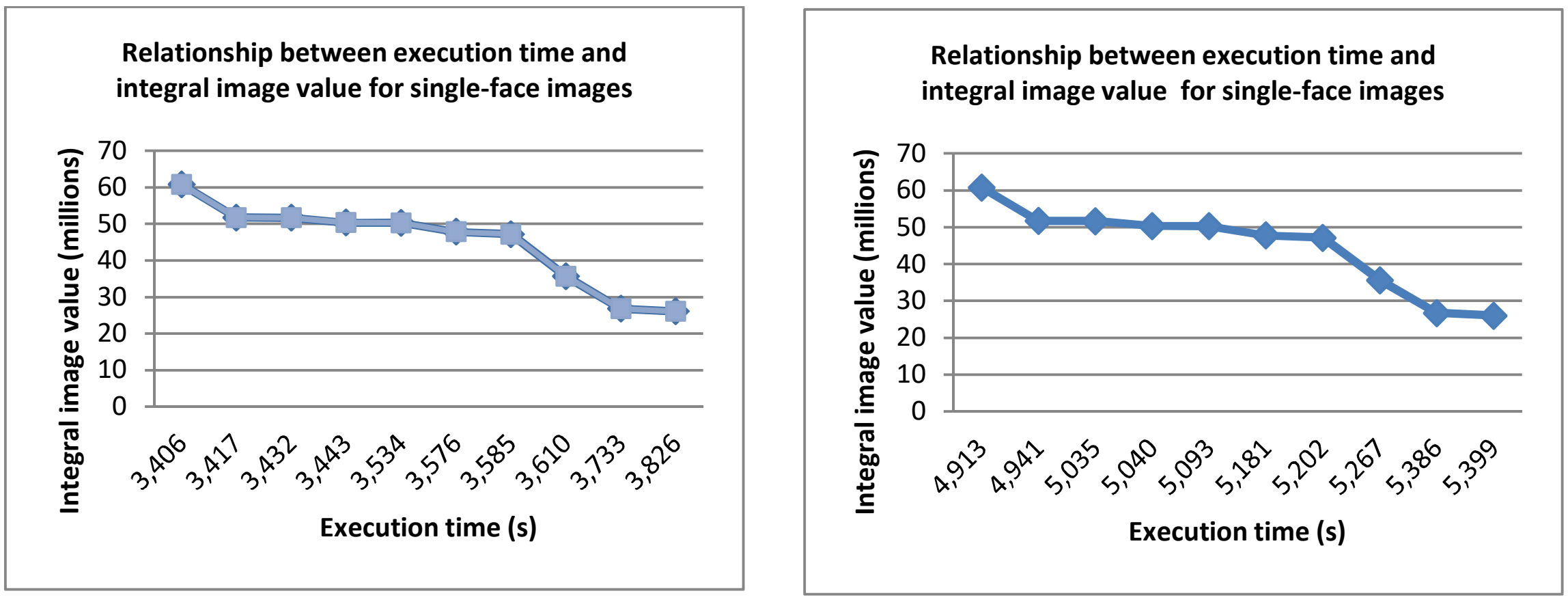

***Figure 11:** Results for **Odroid X4U** (left) and **Raspberry Pi 3 B+** (right) boards.*

As can be seen, as the value of the integral image increases, the execution time decreases, so there is a direct relationship between the number of faces, the value of the integral image and the execution time (RIT), as indicated in Formula 6. Returning to the initial sample with different face numbers, the relationship is shown in Figure 12.

$$RIT = \frac{(\boldsymbol{Execution\ time} \ \cdot \boldsymbol{Integral\ image\ value})}{\textbf{Number of faces in the image}} \quad (6)$$

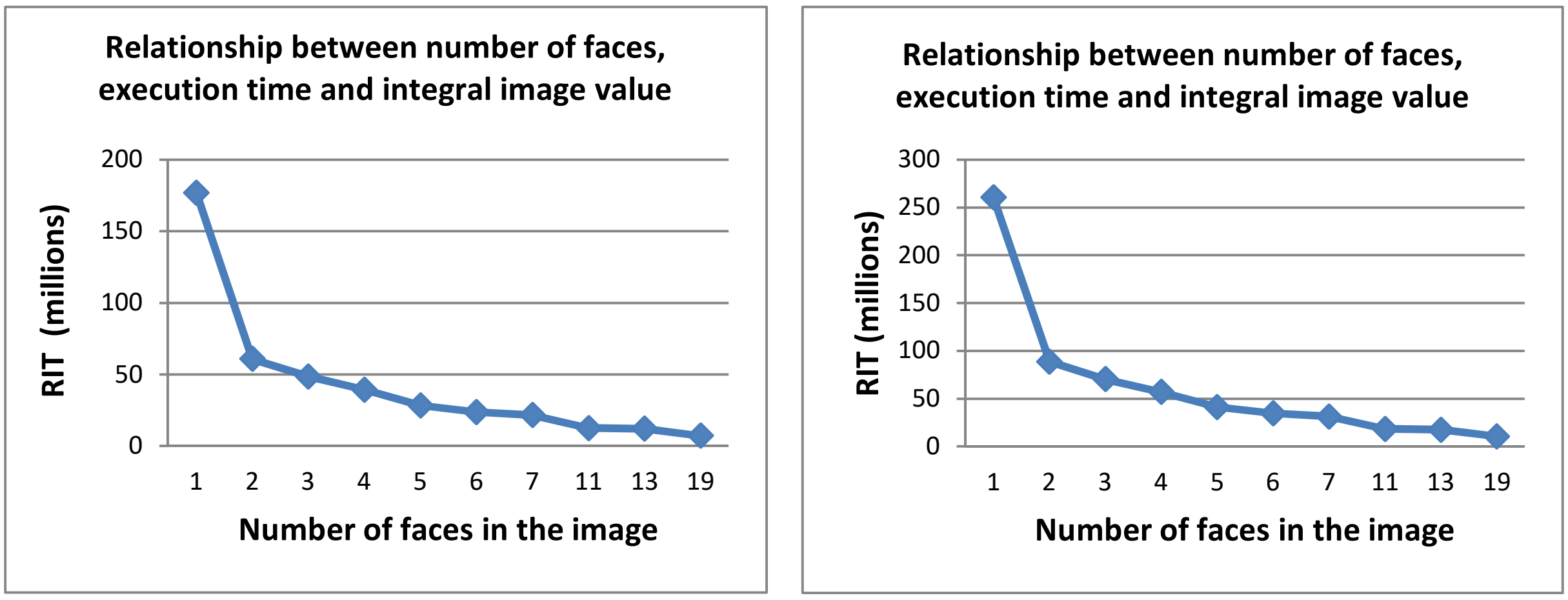

***Figure 12:*** *Results for* ***Odroid X4U*** *(left) and* ***Raspberry Pi 3 B+*** *(right) boards.*

It can be concluded that the execution time is usually affected by the number of faces and the value of the integral image of each specific image. In this case study, images with a lower value for the integral image are usually the ones with higher grey tones.

**6.- Accelerating the execution through parallelization**

Given the above results and, since both case study boards are multi-core, a way to accelerate the program's execution is by parallelizing the program and performing an optimal task scheduling among the existing cores on each board.
For this purpose, we make use of OmpSs, which provides an easy task-based programming model for the parallelization of new and existing C++ codes on multicore platforms with shared memory. Furthermore, it is currently one of the most widely used task-level parallel programming models [5, 6, 7]. Basically, this programming model is based on including directives (#pragmas), as in other parallel programming models, such as OpenMP. These directives are mostly used to annotate certain code blocks in order to inform that those blocks are tasks; that is, basic scheduling units to be used by the available computational resources [23].

With the aim of detecting the most consuming part of the algorithm, profiling tools reports a general overview of any application in a easy way. Profiling tools are often specific for certain compiler toolchains, and sometimes even included with the compiler toolkit itself. Several free alternatives are available for C++ software profiling in GNU environment such as:

1. gperftools – Lightweight and easy-to use tools that report correct results also for multi-threaded parallel algorithms. Clearly the tool of choice for the related OpenMP parallel optimization exercise.
2. gprof – GNU classic that is it's available with gcc compiler tools. Gprof is useful for usual single-core algorithm execution analysis, but it is not thread-safe in embedded system so it reports corrupted results for multi-threaded parallel algorithms. Gprof might yet be used for initial detection of hot-spot routines prior to parallel optimization where other tools are not available.
3. valgrind – In-depth execution analysis tool that works by simulating virtual processor. This approach makes it however very slow in embedded system. The virtual simulation is also done for a single-core processor, therefore Valgrind does not produce realistic profiling figures for multi-thread parallel algorithms.

At this point, we proceed to perform a software profiling to analyze the most time-consuming stages during the sequential execution of the program.

Figure 13 shows the profiling results obtained during the execution of the sequential code on both platforms. For the profiling, the Gperftools software, a profiling tool provided by Google, has been used. Gperftools works by performing time-based sampling, which enables a correct analysis of applications with multiple sub-processes, thus being a suitable tool to obtain the necessary information of which part of the code consumes most of the time, being these parts the first candidates to accelerate.

```
Using local file ./vj.
Using local file vj.prof.
Total: 1018 samples
Time(s) %Execution %Total  Calls     Function
  6.50    63.9%    63.9%  13857361  evalWeakClassifier
  1.98    19.4%    83.3%    862119  runCascadeClassifier
  1.36    13.4%    96.7%    862119  int_sqrt
  0.18     1.8%    98.4%        17  integralImages
  0.06     0.6%    99.0%        17  ScaleImage_Invoker
  0.06     0.6%    99.6%        17  nearestNeighbor
  0.01     0.1%    99.7%         1  munmap
  0.01     0.1%    99.8%         1  partition
  0.01     0.1%    99.9%         1  predicate
  0.01     0.1%   100.0%        17  setImageForCascadeClassifier
```

```
Using local file ./vj.
Using local file vj.prof.
Total: 1940 samples
Time(s) %Execution %Total  Calls     Function
 12.88    66.4%    66.4%  13857361  evalWeakClassifier
  3.64    18.8%    85.2%    862119  runCascadeClassifier
  2.18    11.2%    96.4%    862119  int_sqrt
  0.36     1.9%    98.2%        17  integralImages
  0.14     0.7%    99.0%        17  ScaleImage_Invoker
  0.11     0.6%    99.5%        17  nearestNeighbor
  0.06     0.3%    99.8%        17  setImageForCascadeClassifier
  0.01     0.1%    99.9%         1  munmap
  0.01     0.1%    99.9%         1  partition
  0.01     0.1%   100.0%         1  predicate
```

***Figure 13:*** *Profiling results for* ***Odroid X4U*** *(left) and* ***Raspberry Pi 3 B+*** *(right).*

By studying the results, it can be seen that the largest computational cost in the CPU arises when features are being calculated in each stage (*evalWeakClassifier* function). In the list, the first three functions are the ones which consume more than 96% of the execution time. It is clear that the effort must be focused on optimizing these three functions, since reducing the time for these functions will have a large impact on the execution time of the entire program. The remaining functions represent less than 5% of the total execution time, so are not considered for optimization.

The OmpSs programming model provides simple and non-invasive mechanisms to parallelize programs. It is based on a runtime and a scheduler, which split the code into a set of tasks, identifying the dependencies among them and throwing for execution only ready tasks, that is, those tasks whose dependencies have been satisfied in the different computational cores of the system [24,25].
Once the annotated code has been compiled by Mercurium [6] (the OmpSs compiler), it can be executed using Nanox [16], the OmpSs runtime which consists of a set of libraries to manage the program execution in order to execute it in the most efficient way. Accordingly, OmpSs is suitable for those problems where the same function or instruction set needs to be applied to a different (disjoint) data collection. This is the most frequent problem when it is necessary to execute repetitive instructions or enumerated lists, such as:

```
for (i=start;i<end;i++)
    x[i] = function(a[i])
```

Given that the number of elements to use is known and the operations are usually inside a “for” loop, the OmpSs designers implemented a special operation to be used in the “for” loop, so programmers do not have to change their programs. It is OmpSs that splits indexes and assigns them to the different existing threads.
In the proposed system, the most time-consuming function is "evalWeakclassifier", which finds the value of the weak features which make up the strong classifier. This function is called from "runCascadeClassifier", which is in charge of throwing the cascading classifiers for each one of the 24x24 detection windows that are moving across the entire image, as seen in Figure 14.

```
// step indicates pixel number displacement of filter window by the image
// x2 = margin in width to where the filter window can be moved
// y2 = margin in height up to where the filter window can be moved
for( x = 0; x <= x2; x += step )  //detection window shifting by column
   for( y = y1; y <= y2; y += step) //detection window shifting by row
     {
              p.x = x; //starting coordinates x and y for a 24x24-pixel filter window
              p.y = y;

              result = runCascadeClassifier( cascade, p, 0 );
     }
```

***Figure 14:*** *Source code calling to runCascadeClassifier function.*

After code analysis, and having seen that cascading classifiers are executed every time the detection window is moved across the image, we can conclude that there is an opportunity to optimize the code by parallelizing it using OmpSs. Given the parallelization directives existing in OmpSs that are especially designed for “for” loops, where in each iteration similar tasks are executed that are not independent of each other, that is, each iteration represents a different detection window, it is possible to have several execution threads in parallel. This allows the evaluation of several detection windows at the same time, which improves the execution time compared with the sequential version of the code.

In this study, to optimize the code it is only necessary to use two #pragma directives existing in OmpSs, namely ”#pragma omp for schedule(static)” and “#pragma omp task”. The first one indicates to the compiler that the execution of the loop can be split into several parallel threads, so each thread can be executed on a different CPU core. The second one marks the declaration of a “for” loop as a task, which means that the loop is the only fragment of the code that will be executed in parallel, and once the loop concludes, the remaining code will continue to be executed on a single thread. In Figure 15 we can see the code once the OmpSs directives have been added.

```
#pragma omp for schedule(static)
for( x = 0; x <= x2; x += step )
#pragma omp task
   for( y = y1; y <= y2; y += step )
    {
            p.x = x;
            p.y = y;
            result = runCascadeClassifier( cascade, p, 0 );
    }
```

***Figure 15:*** *Source code for included #pragma directives.*

The cascade classifier can reduce the computational workload by rejecting a region in initial stages, but on the other hand it introduces dependencies between stages that make it difficult to parallelize the program. It is possible to break the dependency between stages by delaying the rejection of a region until the last stage, but that can considerably increase the computational workload. In this context, a balance between parallelism and the optimal computational workload can be achieved, if a static programming model is chosen in which blocks of the same size are processed in parallel (#pragma omp for schedule (static)), together with parallelizable tasks (#pragma omp task), which correspond to each of the features that make up a stage.

After adding these directives, the code is executed again in the same testing environment, under the same conditions. The results obtained are shown in Figure 16, which gives an execution time comparison between the code with and without the OmpSs directives. The execution time is reduced proportionally to the number of cores used in the test.

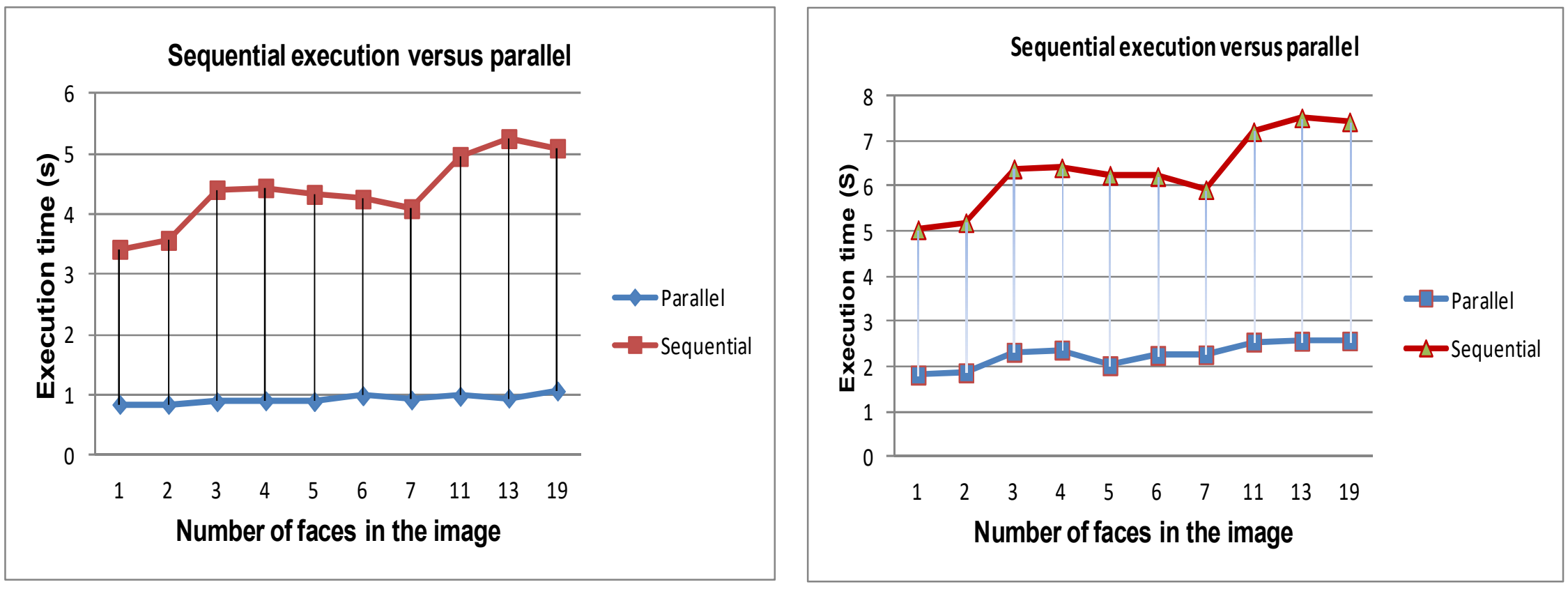

Odroid XU4 | Raspberry Pi

***Figure 16:*** *Sequential and parallel execution time for **Odroid X4U** (left) and **Raspberry Pi 3 B+** (right).*

By studying the above results, and to the contrary of what might be expected, it can be observed that when using the Raspberry Pi 3 B+, with 4 CPU cores, the execution time improvement is nearly 50%, which means that the reduction is not proportional to the number of cores. The reason for this deviation is that executing a division algorithm across four parallel threads leads to an overhead for synchronization tasks, which means that not all the cores can be devoted to the parallel execution of the threads.
In the Odroid XU4 board, the improvements obtained are even worse than expected, if they are compared with the fact that the parallel version is using 8 cores instead of 4. Nevertheless, previous works have proved that using LITTLE cores has not a high impact over the performance when compared with the big ones, even increasing the execution time in some cases [23].

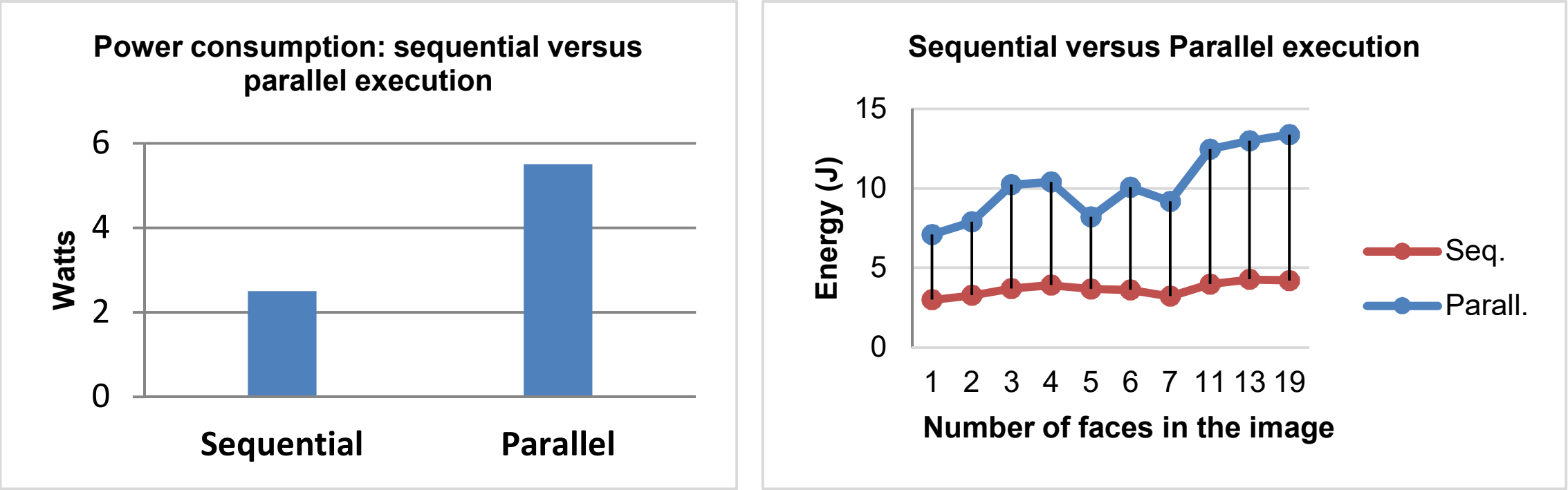

***Figure 17:*** *Energy consumption of sequential and parallel executions before applying any optimization techniques on Raspberry PI 3 B+ platform.*

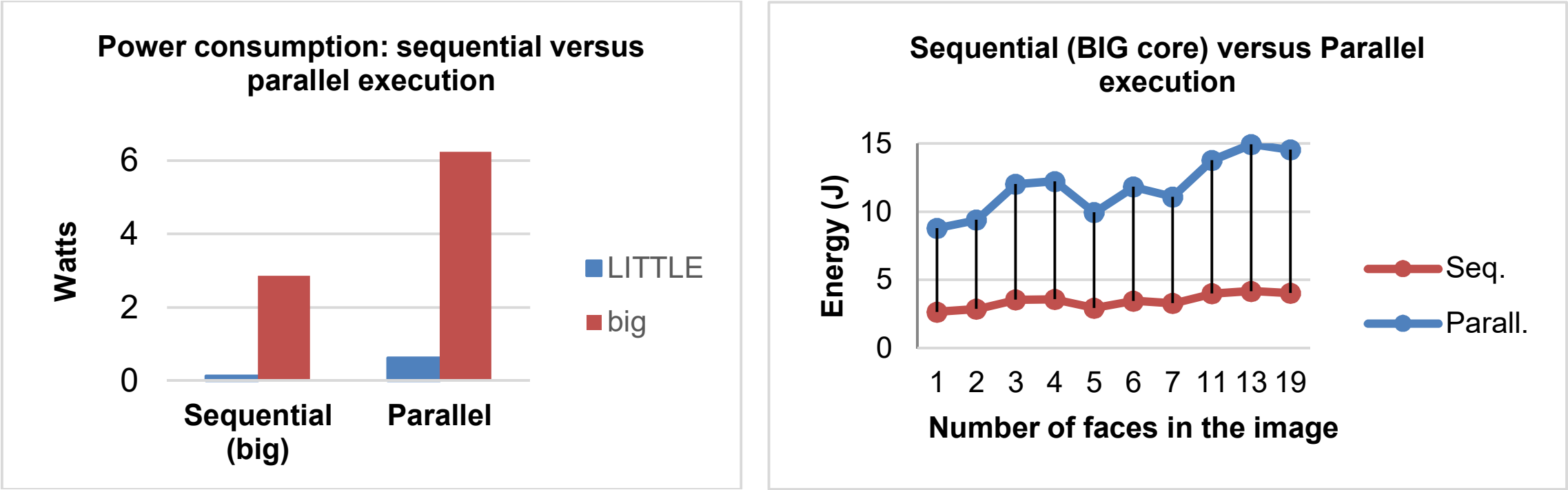

***Figure 18:*** *Energy consumption of sequential and parallel executions before applying any optimization techniques on the Odroid platform.*

The satisfactory results obtained by accelerating the code have one disadvantage, namely the energy consumption.

On the one hand, the power consumption on the Raspberry Pi3B+ platform releases a value of 2.5 Watts and 5.5 Watts for Sequential and Parallel implementations respectively. On the other hand, the power consumption on the Odroid platform shows that the total amount of instant power needed in the sequential execution increases from 3.0 Watts (executed on a big core) up to 6.85 Watts in a parallel execution using all the cores.
This increase in power makes, despite the improvements on the execution time, that the total amount of power (Joules) consumed during the execution increases as shown in the figures 17,18. However, this disadvantage can be overcome by using different optimization techniques described in the next section, as the ones included in the OmpSs runtime (Nanox), generic DVFS *(Dynamic voltage and frequency scaling)* techniques, or specific techniques for asymmetric architectures.

## 7.- Energy consumption optimization

Asymmetric multicore processors with a regular instruction set (ISA) have recently been proposed as a low-cost alternative to conventional symmetric multicore processors since fast and high-performance cores are present together with slower and low-power cores on the same chip, making it possible to optimize the energy consumption in the processor. In this case, as indicated above, the Odroid XU4 [9] board will be used to perform energy consumption optimizations.
Asymmetric architectures allow a reduction in energy consumption by allocating tasks to cores with different characteristics depending on the specific performance and energy consumption requirements of each task. Furthermore, the architecture enables adjustment of performance and consumption on each cluster (that means all the big cores at the same time or all the LITTLE ones) by means of applying Dynamic Voltage and Frequency Scaling (DVFS) mechanisms, which are now available on processors [26].

Next we will see different applicable techniques to improve the performance of the implemented algorithm.

### *7.1.- Optimization of energy efficiency over asymmetric architectures*

The OmpSs runtime: Nanox, is in charge of controlling the execution of the program and trying to finish the execution of it in the most efficient way possible. But with the current growth in asymmetric architectures, OmpSs has recently introduced a new scheduler, namely the bottom level-aware scheduler (Botlev) [27], which is specific to this type of architectures. Botlev is based on traditional schedulers for heterogeneous architectures [28], distinguishing only two kinds of computing nodes (a fast one consisting of Big-type cores, and a slow node for LITTLE-type cores) and removing the cost related to data transferences. One technique used to boost the performance of task-based parallel programs is ensuring that critical tasks end as soon as possible. Given a directed acyclic graph (DAG), a critical task is the one whose delay impacts negatively on the execution time of the entire application. The Botlev scheduler pursues this goal by trying to dynamically determine which tasks belong to the critical path of the DAG associated with the problem, and execute these tasks on fast cores in order to complete its execution as soon as possible.

In order to determine whether a task belongs to a critical path, Botlev sets a priority to each task at the moment of being inserted in the dependency graph, and this is updated when new tasks are created. When a Big core finishes the execution of a task, it will begin with the next critical one in the queue, whereas a LITTLE core will execute the first task in the non-critical task queue [27].

The main difference between the Botlev and the others Nanox´s schedulers, is that Botlev is a scheduler aware of the asymmetry of the architecture which allows it to make dynamic decisions based on the critical path that makes the execution of the program more efficient than with the Nanox conventional scheduler, which is not aware of such asymmetry. For this reason, this scheduler is suitable to be applied to the Odroid XU4 board.

The impact on the energy consumption of task planning will depend on the critical path of the program, and therefore on the critical tasks and the dependences between them. In Figure 19, it can be seen the DAG developed to find restrictions and dependencies in order to optimize the source code for an adequate parallelization and energy consumption.

Each stage is dependent on each other because an image can be rejected in one stage and it would not be necessary to calculate the next pending stages of the cascade classifier. That is, if each stage is executed in parallel tasks, steps that are not necessary could be calculated. However, examining the calculation of characteristics of a stage, at first glance it could be considered to be independent.
However, there is a shared variable so-called "stage_sum" that gathers the values of the output classifier features to be compared with the aforementioned threshold that determines the chance of being a face.

This share variable produces dependence since it is calculated sequentially to obtain the accumulated sum of each one of the characteristics of the stage that is being evaluated. To avoid this dependency and in order to limit the parallelization of the software, this variable can be split into parts by using an array that contains both elements and threads.

### *7.2.- DVFS on big-LITTLE architecture*

Another type of technique for improving energy consumption corresponds to those based on the DVFS (dynamic voltage frequency scaling) techniques [29], which are characterized by dynamically varying the processor's frequency and voltage at execution time in order to improve energy consumption or reduce the instant power consumed. Big.LITTLE ARM processors support frequency scaling, although only at cluster level, not at core level, so every core in a cluster will execute tasks at the same frequency [17].

Access to this feature of the processors in Linux kernel is done through the subsystem "cpufreq", which provides the library libcpufreq and executables cpufreq-set and cpufreq-info that allow to obtain and modify the frequency of each kernel in execution time. This frequency management will allow us to contain the energy expenditure and therefore we will save energy during the execution of the program.

A plausible solution to save energy is to take advantage of the periods of time in which the workload on the slow core diminishes in order to reduce the energy consumption of them, forcing a reduction in frequency over the LITTLE cluster. Reducing the frequency in the LITTLE cluster implies that the instantaneous power dissipated decreases. Although it is true that by reducing the frequency of the cluster, the time spent executing a task increases, this technique can be applied in those phases of parallel execution in which the execution is limited by a large number of critical tasks and a low number of non-critical tasks; therefore, it is expected that the final impact on performance will not be high, and so will the reduction in energy consumption [17].

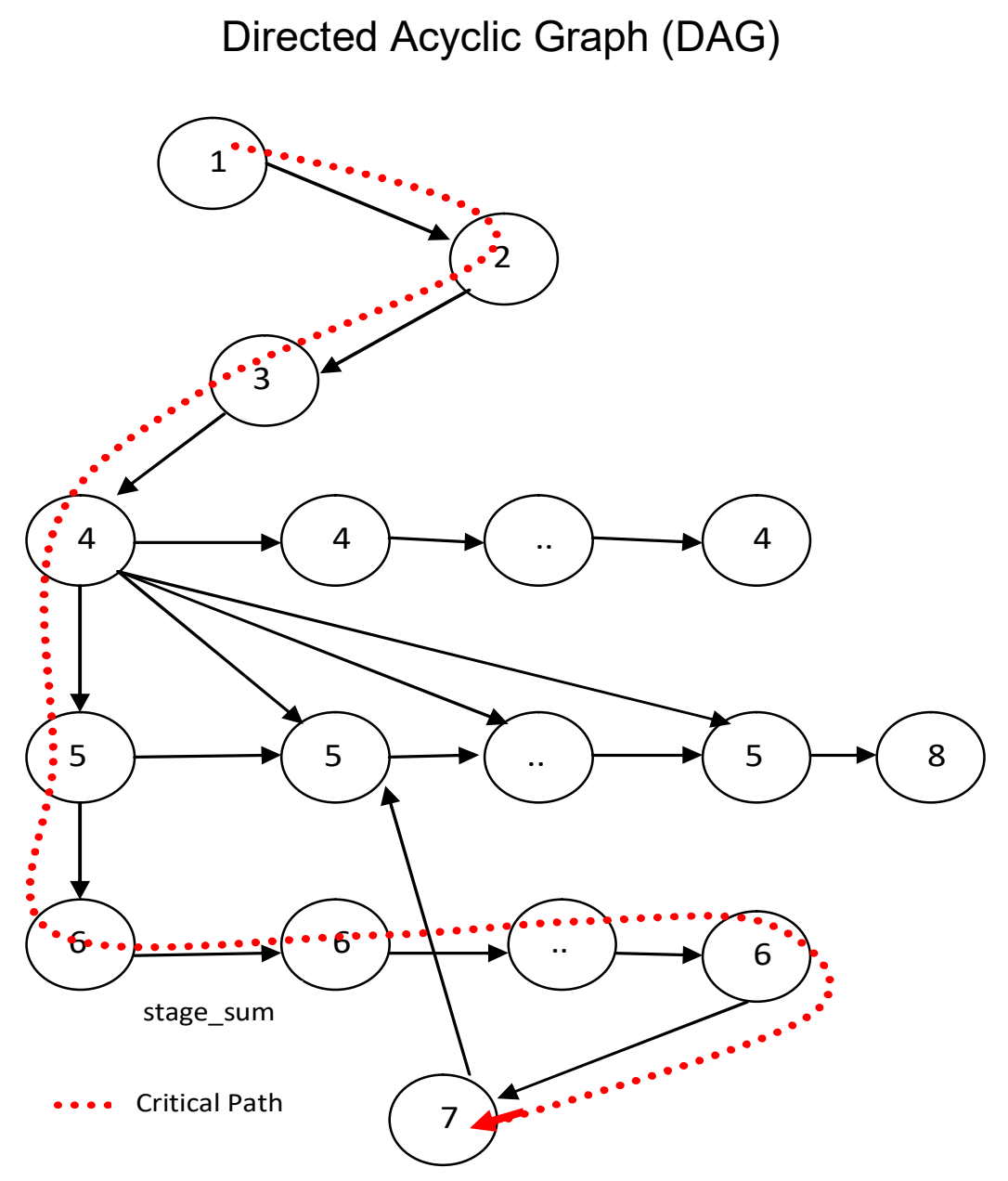

Task DAG processes of the face detection program:

1. Start, image processing.
2. Image scaling
3. Integral image calculation
4. Scrolling window detection by the image
5. Analysis stages cascade classifier
6. Stage feature Calculation
7. Stage threshold comparison
8. Analysis result

*Figure 19: Directed Acyclic Graph (DAG) of the face detection program.*

### *7.3.- Optimization of consumption based on the parameters step and scaleFactor*

Finally, as indicated above, there are two configurable software parameters whose variation can reduce the execution time and the energy consumption, at the cost of reducing the face detection rate. These parameters are the scale factor (which determines how the image resolution decreases in each iteration), and the step factor, (which determines how the detection window slides across the image in each iteration).

The "Step" and "scaleFactor" parameters can greatly influence the accuracy of detection. For this reason, and in order to find the optimal parameters for facial detection, it has been analyzed the effect of these two parameters on the ability to detect faces (measured in the number of faces detected, the number of false positives, and the number of false negatives). In Figure 20, it can be seen the evolution of the total detection error (false negatives + false positives) based on the different values of each parameter.
Regarding experimental stimuli, two public databases were used, Base-750 [30], and Base-450 [31]. The first one contains 750 images (480x640 resolution) whereas the second one has 450 images (896x592 resolution). Both have one face per image.

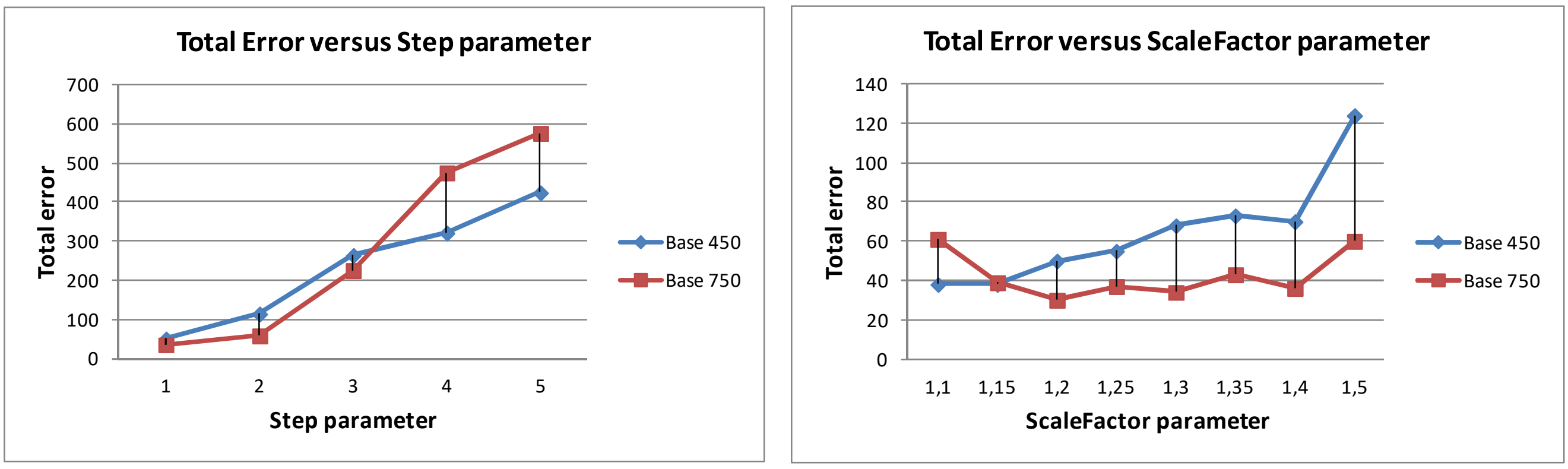

***Figure* 20:** *Evolution Total error based on the parameters Step and ScaleFactor in the two bases of public faces: Base-450, Base-750.*

From the results obtained we can highlight the following conclusions:

- The “step” parameter is more sensitive than "scaleFactor", on the accuracy of the program, any value greater than 2 produces a large increase in detection errors, being the optimal value for detection 1. However, it occurs a considerable improvement of the program performance since it considerably reduces the execution time and energy consumption by considerably reducing the number of operations. Therefore, for a step equal to 2, although some accuracy is lost due to the increase of the total error, its consideration may be of interest when the objective is to increase the performance and the decrease of the energy consumption.

- The "scaleFactor" parameter, in contrast to the "step" parameter, is less sensitive in terms of accuracy. The total error increases slowly as the value of this parameter increases. It also increases the performance of the program but on a smaller scale than with the "step" parameter. In this context, the value that provides the best performance and causes lower energy consumption could be evaluated as the optimal value for this parameter.

### *7.4.- Evaluation of energy Improvements*

In this context, taking into account the scheduler provided by OmpSs, the scaling of frequencies, and the variation of the scale and step factors, a study has been carried out for the Odroid XU4 board in order to optimize the energy consumption considering different scenarios. For the experiments, only the frequency of the big cluster has been modified, because modifying the frequency of the LITTLE cluster has not a meaningful impact on the energy consumption, but a big impact on the execution time as shown in [17]. Figures 21-24, show the obtained results for different frequency values on cluster BIG (Odroid XU4 board) after processing 1200 images included in both databases (Base-450 and Base-750).

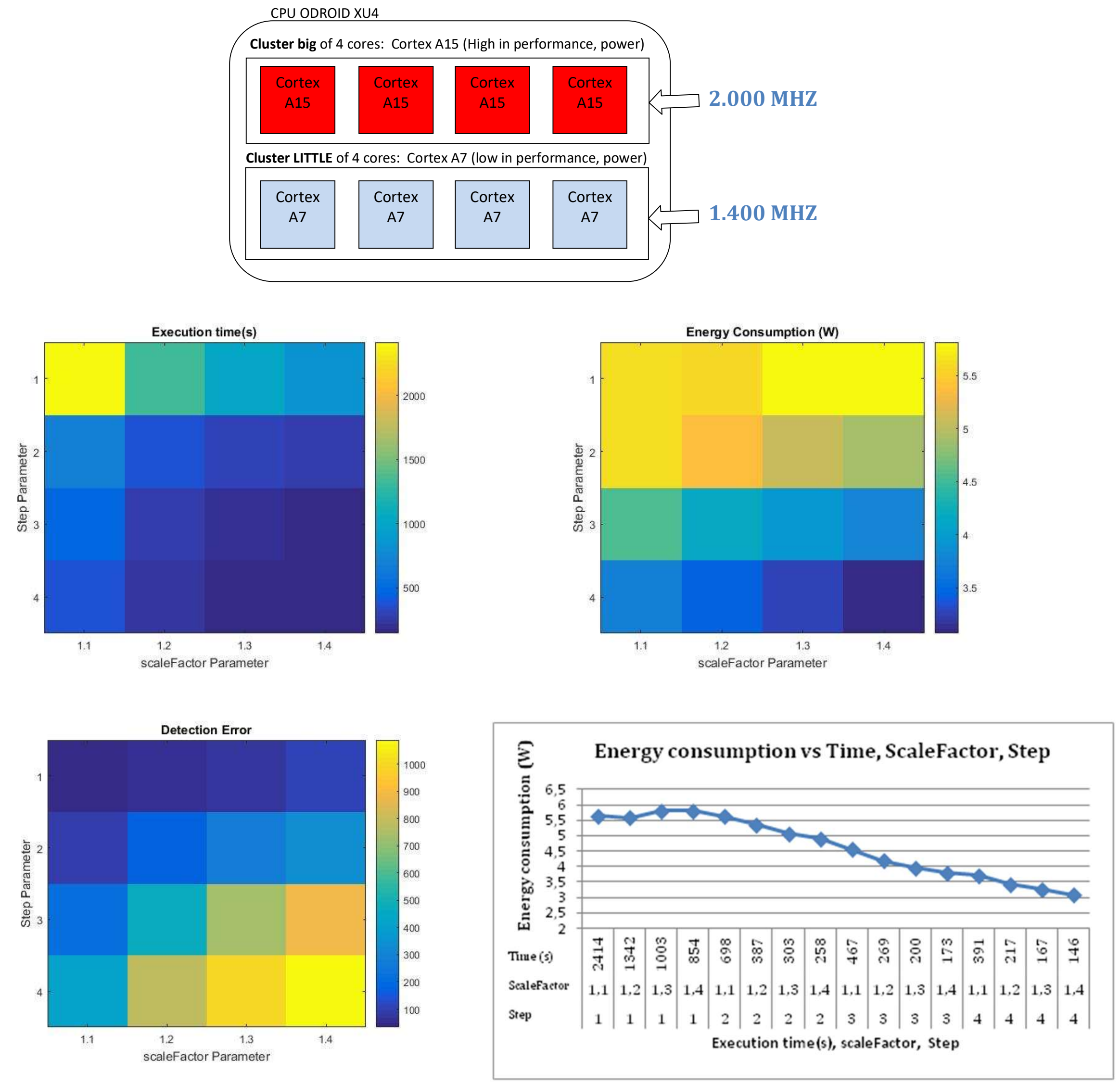

***Figure 21***: *Execution time, energy consumption, and detection error according to the parameters "scaleFactor" and "step", and the impact of execution energy consumption vs Execution time, "scaleFactor" and "step" parameters, for the cluster frequencies: Big = 2000 MHz, LITTLE = 1400 MHz.*

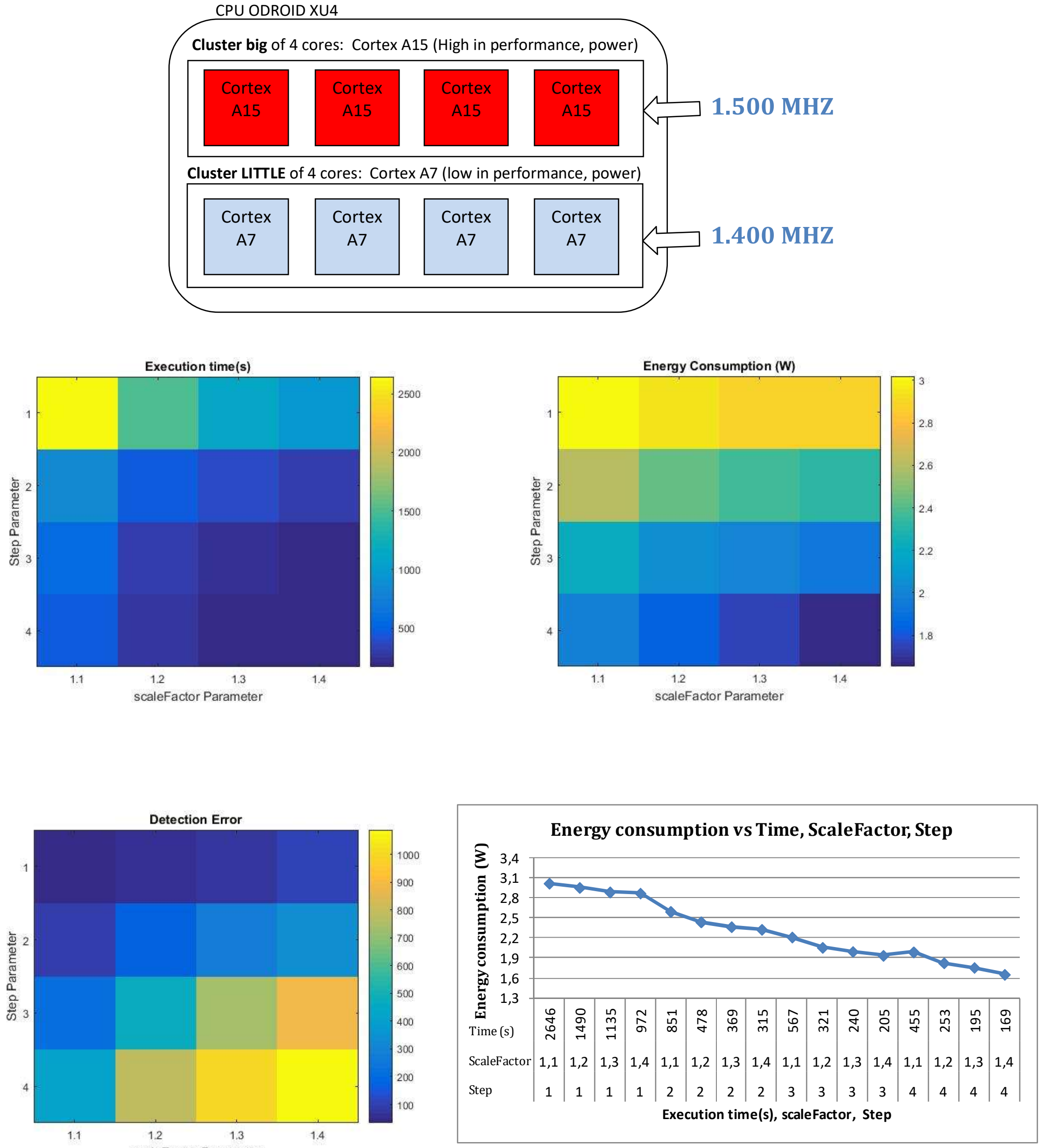

***Figure 22****: Execution time, energy consumption, and detection error according to the parameters "scaleFactor" and "step", and the impact of execution energy consumption vs Execution time, "scaleFactor" and "step", parameters for the cluster frequencies: Big = 1500 MHz, LITTLE = 1400MHz.*

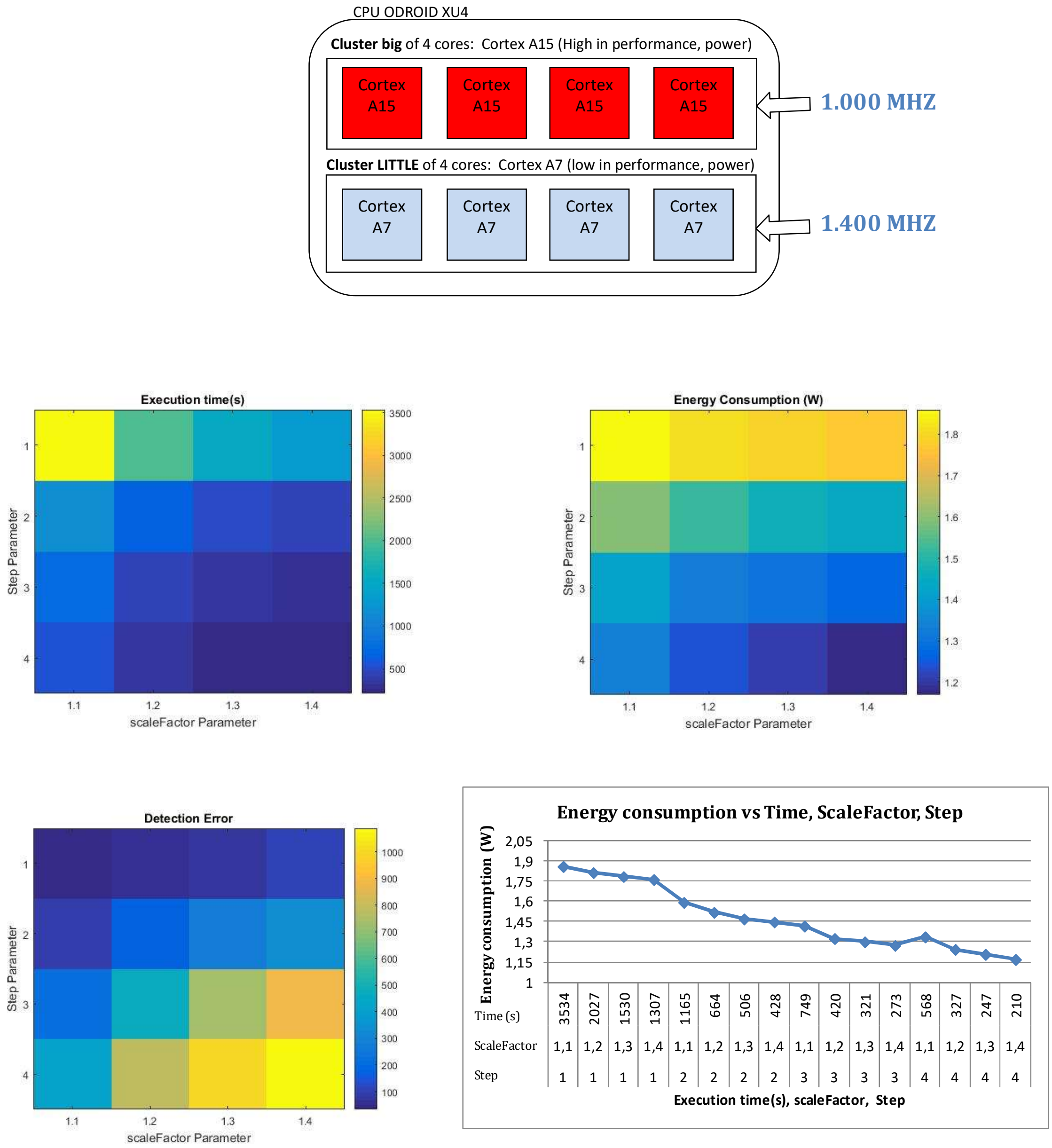

***Figure 23**: Execution time, energy consumption, and detection error according to the parameters "scaleFactor" and "step", and the impact of execution energy consumption vs Execution time, “scaleFactor” and “step”, parameters for the cluster frequencies: Big = 1000 MHz, LITTLE = 1400MHz.*

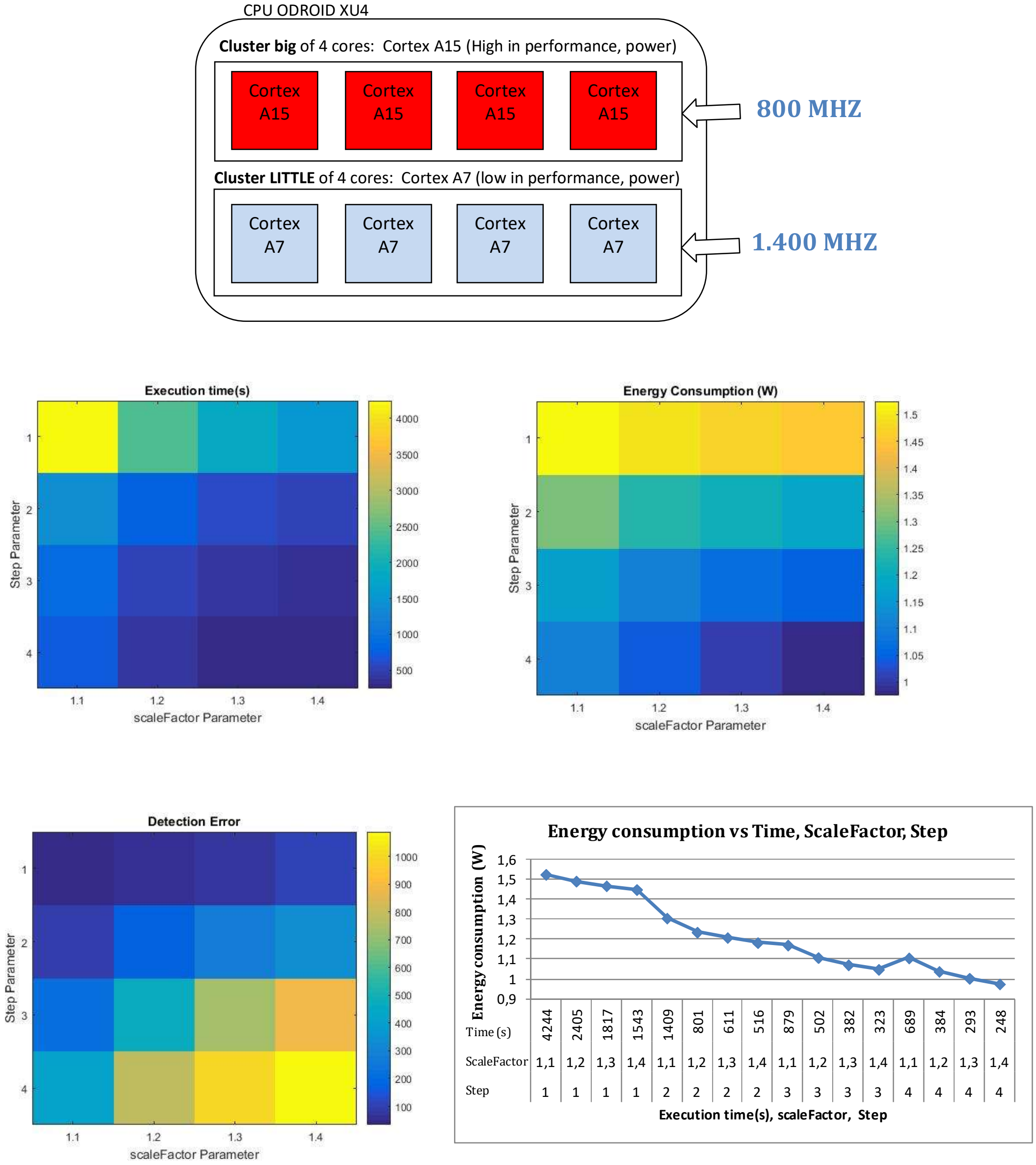

*__Figure 24__: Execution time, energy consumption, and detection error according to the parameters "scaleFactor" and "step", and the impact of execution energy consumption vs Execution time, "scaleFactor" and "step", parameters for the cluster frequencies: Big = 800 MHz, LITTLE = 1400MHz.*

As shown in all the plots, the *"step"* parameter has a high impact on the energy consumption, but also in the number of faces detected as shown previously. When dealing with embedded and mobile systems, not only the execution time nor the energy consumption is important on their own, but the trade-off between them. Whit this purpose, Table I shows the optimal values found in order to tolerate an error constraint less than 10% of the total faces with the best detection time and the lowest possible energy consumption.

The application of these parameters to the face detection program produces an improvement in the execution time of around 50% in the Raspberry Pi 3 B+ board and 65% for the Odroid XU4 board, with respect to the sequential execution time.

***Table I.*** *Optimal values to reduce energy consumption and accelerate execution time for a 90% detection ratio.*

| Cluster Big Frequency (Odroid XU4) | Cluster LITTLE Frequency (Odroid XU4) | Step Parameter | ScaleFactor Parameter |
|---|---|---|---|
| 1500 MHz | 1400 MHz | 1 | 1,2 |

Regarding the optimization of energy consumption, an environmental reduction of 22.3% is achieved with respect to the sequential execution of the Odroid XU4 board. This is due to the frequency scaling and the use of the resources provided by the asymmetric architecture and exposed through the OmpSs task scheduler. In this case, setting the frequency of the big cluster to 1500 MHz and that of the LITTLE cluster to 1400 MHz.

For the sake of generality and clarity, we have compared the performance obtained against other well-known developments for face detection (based on OpenCV), which use the function "detectMultiScale", based again on the algorithm of Viola-Jones. In Table II, we can see the result of applying both developments to the two databases used in the current work.

***Table II.*** *Results of applying the OpenCV function: detectMultiScale, and the detection system with selected parameters, to the two databases: Base-450, Base-750.*

| **OpenCV *detectmultiScale* Function** | | | | **Our Detection System with selected parameters** | | | |
|---|---|---|---|---|---|---|---|
| **Base-450** | | | | **Base-450** | | | |
| **False Positive** | **False Negative** | **Execution time (min)** | **Total Error** | **False Positive** | **False Negative** | **Execution time (min)** | **Total Error** |
| 151 | 3 | 29.18 | 154 | 9 | 29 | 18.28 | 38 |
| **Base-750** | | | | **Base-750** | | | |
| 33 | 3 | 24.23 | 36 | 13 | 17 | 13.48 | 30 |

From the results it can be seen how the detection system proposed in the present work improves both the execution time and the total number of errors that occur in the detection of faces with respect to those obtained by the OpenCV function: detectMultiScale.

In pattern recognition, precision (also so-called positive predictive value) is the fraction of relevant instances among the retrieved ones, while recall (or sensitivity) is the fraction of relevant instances that are detected. Both precision and recall are based on an understanding and measure of relevance as shown at Equation 7.

$$\text{Precision} = \frac{\text{True positive}}{\text{True positive+Fal} \quad \text{Positive}} \qquad \text{Recall} = \frac{\text{True positive}}{\text{True positive+False negative}} \quad (7)$$

***Table III.*** *Results of precision and recall for the OpenCV* detectMultiScale *function and the detection system in the experimental bases: base 450 and base 750.*

| | **OpenCV detectMultiScale function** | | **Our Detection System** | |
|---|---|---|---|---|
| | **Base-450** | **Base-750** | **Base-450** | **Base-750** |
| **Precision** | 74,71% | 95,76% | 97,91% | 98,26% |
| **Recall** | 99,33% | 99,60% | 93,56% | 97,73% |

From the results it can also be concluded the detection system holds a higher accuracy rate than its counterpart on OpenCV (function detectMultiScale).

## 8.- Conclusions

In the present work it has been developed a facial detection system based on the Viola-Jones algorithm, 37% faster and more energy efficient than other algorithms of the same type currently available, such as the one provided by the OpenCV library through the function "detectMultiScale", widely used by developers for face detection. It has also been adapted to low-cost embedded devices such as the Raspberry Pi 3 B+ and the ODROID boards, whose characteristics are similar to nowadays smartphone.

For this, a thorough study of the Viola-Jones algorithm has been carried out, analyzing each of its parts in order to find out possible improvements. In this context, it has been possible to obtain a direct relationship between the integral image of an image and the detection speed, which allows us to determine which are the best resolution and the tone of the images to be processed, so that the performance of the face detection system is optimal.

For the acceleration of the detection program, the multi-core architectures of the two experimental boards were used, adapting the program to an efficient parallelization through OmpSs in the first instance. As a result, a reduction in the execution time of the program has been obtained, which ranges from 50% for the Raspberry PI 3 B+ board, and 65% for the ODROID XU4 board.

Due to the increase in the energy consumption associated with the parallelization of the program, different options have been studied to optimize this consumption: (1) take advantage of the resources provided by the asymmetric architecture of the ODROID XU4 board, based on the OmpSs task scheduler and aware of the asymmetry; (2) make use of frequency scaling techniques applied to that board; (3) optimal selection of "Step" and "ScaleFactor" system parameters. With all this, a reduction in consumption of around 24.3% has been achieved with respect to the sequential execution. With this fact, it has also been possible to leverage the capabilities of asymmetric multi-core architectures (ODROID XU4) versus symmetric architectures (Raspberry Pi 3 B+) for the optimization of energy consumption. The parallel execution of our system, using the optimal parameters, achieves a reduction of energy consumption of 21.3% in the Odroid XU4 board with respect to the Raspberry PI 3 B +.

As future work, carrying out a detailed analysis of the characteristics of the cascade classifiers used in face detection can be extremely interesting to improve the execution time and energy consumption in any device. This can be done through the inclusion of new more defining features in the detection process, or by improving Adaboost, the learning method used in the Viola-Jones algorithm.

## 9.- Acknowledgments

This work has been partially funded by EU FEDER and Spanish MINECO research projects TIN 2015-65277-R, FPU15/02050, as well as the UCM-Banco Santander Grant PR26-16/20B-1.

## 10.- Bibliography